\documentclass[amsmath, amssymb, superscriptaddress, reprint, twocolumn, colorlinks=true, citecolor=blue, linkcolor=blue, urlcolor=blue]{revtex4-1}
\usepackage{natbib}
\usepackage[utf8]{inputenc}
\usepackage[T1]{fontenc}
\usepackage{graphicx}
\usepackage{grffile}
\usepackage{longtable}
\usepackage{wrapfig}
\usepackage{rotating}
\usepackage[normalem]{ulem}
\usepackage{amsmath}
\usepackage{textcomp}
\usepackage{amssymb}
\usepackage{capt-of}
\usepackage{hyperref}
\usepackage{mathptmx}
\usepackage{upgreek}
\usepackage{float}
\usepackage{bm}
\usepackage{booktabs}
\usepackage{subfigure}
\usepackage{url}
\usepackage{tablefootnote}

\date{\today; Authors to whom correspondence should be addressed: fuyangyang@tsinghua.edu.cn}

\title{}

\begin{document}

\title{Spatial Fluctuation of the Electric Field within SF\(_6\) Streamer Channel in Highly Non-Uniform Fields: Phenomenon, Validation, and Mechanism}

\author{Zihao Feng}
\affiliation{Department of Electrical Engineering, Tsinghua University, Beijing 100084, China}
\author{Xinxin Wang}
\affiliation{Department of Electrical Engineering, Tsinghua University, Beijing 100084, China}
\author{Xiaobing Zou}
\affiliation{Department of Electrical Engineering, Tsinghua University, Beijing 100084, China}
\author{Haiyun Luo}
\affiliation{Department of Electrical Engineering, Tsinghua University, Beijing 100084, China}
\author{Yangyang Fu*}
\affiliation{Department of Electrical Engineering, Tsinghua University, Beijing 100084, China}
\affiliation{State Key Laboratory of Power System Operation and Control, Department of Electrical Engineering, Tsinghua University, Beijing 100084, China}

\begin{abstract}

The electric field within the streamer channel is a critical parameter in the calculation model for the nonlinear breakdown voltage of SF\(_6\), motivating the research presented in this paper. By using a 2D fluid model, we investigate the microscopic characteristics of the SF\(_6\) streamer channel in highly non-uniform fields and uncover a previously unexplained coherent structure: the spatial fluctuation of the electric field (SFEF). We validate the physical validity of SFEF by modifying model parameters that could potentially introduce non-physical effects. Further comparative analysis reveals that SFEF is driven by an ion-conducting channel formed due to the strong electronegativity of SF\(_6\). This ion-conducting channel exhibits local characteristics, which fundamentally arise from the slow response of charged species to local charge relaxation. We identify that some charge separation originates from the accumulation of negative ions at the rear edge of the streamer head due to strong electric field shielding in this region. As the streamer propagates, charge separation is continuously generated and passively carried into the streamer channel, ultimately forming the SFEF. Finally, we confirm that SFEF does not occur in uniform fields, indicating that it is a phenomenon exclusive to highly non-uniform fields. These findings provide a deep insight into the electric field within the SF\(_6\) streamer channel and offer a potential avenue for further investigation into the mechanisms of SF\(_6\) nonlinear breakdown voltage.

\end{abstract}

\maketitle
\section{\label{1}introduction}

SF\(_6\) is widely utilized as an electrical insulating gas due to its high electronegativity. In recent years, the application of gas-insulated electrical equipment has increased. This has highlighted the breakdown of SF\(_6\) in highly non-uniform electric fields, often induced by contaminants such as metal particles with protrusions, as one of the weakest insulation links in the equipment \cite{10007898,10089517}. Previous research has extensively examined the breakdown characteristics of SF\(_6\), emphasizing its special nonlinear voltage behavior \cite{Waters_2019,Seeger_2014,ffded,Seeger_2009,12321}. This nonlinear behavior is primarily governed by two decisive processes: streamer and leader. Recent research has proposed calculation models for both processes, including the \textit{SF\(_6\) streamer breakdown calculation method} proposed by Wu \textit{et al.} \cite{Wu_2021} and the \textit{SF\(_6\) streamer to leader model} proposed by Seeger \textit{et al.} \cite{Seeger_2014}. The motivation of this paper arises from the fact that the above mentioned models employ approximate assumptions for the electric field strength within the SF\(_6\) streamer channel, which may not fully reflect the influence of space charge on the channel field. These assumptions have been proven to be applicable in uniform fields \cite{PhysRevA.35.1778}. However, recent experimental research have indicated that the SF\(_6\) streamer channel in highly non-uniform fields exhibited unusual characteristics \cite{Wu_2019,https://doi.org/10.1049/hve2.12119,Zhao_2022}, yet the microscopic characteristics of the SF\(_6\) streamer channel are still not fully understood.

For the nonlinear dynamic process of high-pressure streamer discharge, the self-consistent fluid simulations can resolve the spatiotemporal microscopic characteristics \cite{Nijdam_2020}. Although numerous 1D fluid simulations of SF\(_6\) discharge have been conducted \cite{10.1063/1.4892637,106801,85110,SF62002,8551279,10.1063/5.0008411}, they cannot fully capture the spatial local effects in highly non-uniform fields. Some 2D fluid simulations of pure SF\(_6\) streamer discharge have also been conducted, Gallimberti \textit{et al.} \cite{SF6111} and Francisco \textit{et al.} \cite{Francisco_2021} artificially set the streamer channel to be electrically neutral in their models, thereby ignoring the dynamic processes within the channel. Dhali \textit{et al.} \cite{14} , Wang \textit{et al.} \cite{GDYJ200807009} and Li \textit{et al.} \cite{GDYJ201410015} employed the flux-corrected transport (FCT) method to simulate SF\(_6\) streamer in uniform fields, but the channel characteristics was not discussed in their paper. Furthermore, existing simulations \cite{PhysRevA.35.1778, 10.1063/5.0223522} examined a limited range of discharge parameters, making it insufficient for a comprehensive analysis of the general physics in the SF\(_6\) streamer channel. To date, the dynamics of SF\(_6\) streamer channel remain insufficiently understood, hindering a deeper insight into the underlying mechanism of nonlinear breakdown characteristics. Consequently, more comprehensive and in-depth investigations into the dynamic processes in SF\(_6\) streamer channel in highly non-uniform fields is still awaiting.

In this paper, we employ a 2D axisymmetric fluid model to investigate the SF\(_6\) streamer channel, with a detailed description of the model provided in Section \ref{2}. In Section \ref{3}, through comprehensive studies of key discharge parameters, we identify general coherent structures within SF\(_6\) streamer channel, that differ from those in air. These coherent structures arise from the nonlinear dynamics of SF\(_6\), driven by its strong electronegativity. In all subsequent sections of the paper, we focus on a  previously unexplained coherent structure: spatial fluctuation of the electric field (SFEF).

Specifically, in section \ref{4} we validate the physical validity of SFEF, by modifying two key factors within our model—photoionization intensity and mesh size—that could introduce potential non-physical effects. The validation is intended to rule out the possibility that these factors are causing SFEF to occur as a non-physical phenomenon. In section \ref{5.1} and \ref{5.2} we investigate the underlying physical mechanism dominating the formation of SFEF, by adjusting the applied voltage to the overvoltage level—an idealized scenario that exists only theoretically—we attempt to identify conditions under which the SFEF phenomenon becomes less pronounced. Comparative analyses are then conducted to determine the essential factor governing SFEF and to reveal the underlying physical mechanism driving it. In Section \ref{5.3}, we investigate the spatiotemporal evolution of fluctuation during its initial stage, aiming to reveal the physical mechanism responsible for the transition of SFEF from its absence to occurrence. Finally in Section \ref{6}, we simulate SF\(_6\) streamers in a uniform background electric field in order to investigate whether SFEF is a phenomenon exclusive to highly non-uniform fields.

\section{\label{2}Model Description}
\subsection{2D Axisymmetric Geometry}
As shown in FIG. \ref{fig.geo}, to simulate discharge in a highly non-uniform field, which is often caused by millimeter-sized linear metal particles in gas-insulated electrical equipment \cite{8928273}, the high-voltage (HV) electrode is modeled as an elongated rod with a rounded tip. The rod electrode is 5 mm in length, with a curvature radius of 0.1 mm at the tip. The gas gap distance between the rod tip and the ground electrode is denoted as \(d\), varies according to the specific studies discussed below. The entire computational domain has a horizontal size of 25 mm, which is sufficiently large to ensure that edge effects exert a negligible influence on the local electric field in the discharge region. The critical details of the computational mesh in the discharge region are discussed in Section \ref{4.2}.

\subsection{Governing Equations}
The SF\(_6\) streamer consists of electrons, positive ions, and negative ions as the three principal charged species, forming a conductive fluid \cite{15}. The number density of each charged species is described using continuity equations (\ref{Eq1})-(\ref{Eq3}): 

\begin{figure}[t]
\centering
\includegraphics[width=8.5cm]{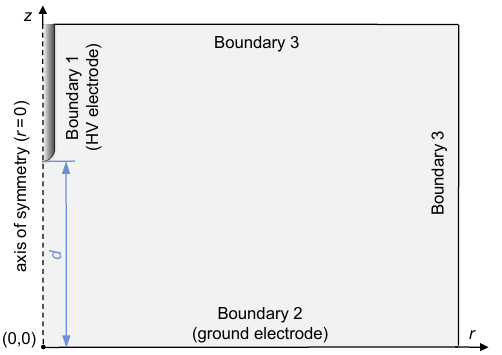}
\caption{\label{fig.geo} Geometry of the 2D axisymmetric simulation domain and the computational boundaries.}
\end{figure}

\begin{table*}[t]
\caption{\label{tab:table1}Expressions for the mobilities \(\mu_\text{i}\) (i=e, p, n) and diffusion coefficients \(D_\text{i}\) (i=e, p, n), in which \(N\) denotes the number density of neutral species with a value of \(2.45×10^{25}\) \(\text{m}^{-3}\) at 1 atm, and \(E/N\) denotes the reduced electric field with the unit in V·\(\text{m}^2\)}

\begin{ruledtabular}
\begin{tabular}{ccccc}
  &\(\mu_\text{i}\) [\(\text{m}^{2}\cdot\text{V}^{-1}\cdot\text{s}^{-1}\)]&\(D_\text{i}\) [\(\text{m}^{2}\cdot\text{s}^{-1}\)]\\ \hline
 electron, i=e&$\begin{aligned}
\mu_{\mathrm{e}} = (1.027 \times 10^{19} (E / N)^{-0.2576})/N \\, 1 \times 10^{-20} < E / N <2 \times 10^{-18} \end{aligned}$ & $\begin{aligned}
& D_{\mathrm{e}}=\left(8.8823 \times 10^{28}(E / N)^{0.2424}\right) / N \\
& ,E / N<6.5 \times 10^{-19}
\end{aligned}$ \\ 
 \\
 positive ion, i=p& $\mu_{\mathrm{p}}=\left\{\begin{array}{l}
6 \times 10^{-5}, E / N<1.2 \times 10^{-19} \\
\\
1.216 \times 10^{-5} \ln (E / N)+5.89 \times 10^{-4} \\
, 1.2 \times 10^{-19}<E / N<3.5 \times 10^{-19} \\
\\
-1.897 \times 10^{-5} \ln (E / N)-7.346 \times 10^{-4} \\
, E / N>3.5 \times 10^{-19}
\end{array}\right.$
 & $D_{\mathrm{p}}=\frac{k_{\mathrm{B}} T_{\mathrm{p}} \mu_{\mathrm{p}}}{e}$\footnote{$k_{\mathrm{B}}$ is the Boltzmann constant, and \(T_{\mathrm{p}}\) is temperature of the positive ion, which is assumed to be equal to \(T_\text{g}\)}\\
 \\
 negative ion, i=n&$\begin{aligned}
\mu_{\mathrm{n}}= &1.69 \times 10^{32}(E / N)^2+5.3 \times 10^{-5} \\&, E / N<5 \times 10^{-19}
\end{aligned}$ & $D_{\mathrm{n}}=\frac{\mu_{\mathrm{n}} k_{\mathrm{T}}}{39.6}$\footnote{\(k_{\mathrm{T}}\) is the Townsend's energy factor, which can be obtained from Ref.\cite{EN3602}} 
 \\
\end{tabular}
\end{ruledtabular}

\end{table*}

\begin{equation}
   \label{Eq1}
\frac{\partial}{\partial t}\left(n_{\mathrm{e}}\right)+\nabla \cdot \boldsymbol{\Gamma}_{\mathrm{e}}=S_{\mathrm{R}, \mathrm{e}}+S_{\mathrm{ph}}
\end{equation}
\begin{equation}
   \label{Eq2}
\frac{\partial}{\partial t}\left(n_{\mathrm{p}}\right)+\nabla \cdot \boldsymbol{\Gamma}_{\mathrm{p}}=S_{\mathrm{R}, \mathrm{p}}+S_{\mathrm{ph}}
\end{equation}
\begin{equation}
   \label{Eq3}
\frac{\partial}{\partial t}\left(n_{\mathrm{n}}\right)+\nabla \cdot \boldsymbol{\Gamma}_{\mathrm{n}}=S_{\mathrm{R}, \mathrm{n}}
\end{equation}
, in which \(n_\text{e}\), \(n_\text{p}\) and \(n_\text{n}\) denote the number densitis of electrons, positive ions, and negative ions, respectively. \(\boldsymbol{\Gamma}_\text{e}\), \(\boldsymbol{\Gamma}_\text{p}\) and \(\boldsymbol{\Gamma}_\text{n}\) denote the fluxes of electrons, positive ions, and negative ions, respectively. \(S_{\mathrm{R}, \mathrm{e}}\), \(S_{\mathrm{R}, \mathrm{p}}\) and \(S_{\mathrm{R}, \mathrm{n}}\) denote the reaction source terms for electrons, positive ions, and negative ions. \(S_{\mathrm{ph}}\) denotes the photoionization source term.

The fluxes include contributions from both drift (dominated by the electric field \(\boldsymbol{E}\)) and diffusion (dominated by the density gradient \(\nabla n\)) , which can be further expressed as:
\begin{equation}
   \label{Eq4}
\boldsymbol{\Gamma}_{\mathrm{e}}=-n_{\mathrm{e}}\left(\mu_{\mathrm{e}} \boldsymbol{E}\right)-D_{\mathrm{e}} \nabla n_{\mathrm{e}}
\end{equation}
\begin{equation}
   \label{Eq5}
\boldsymbol{\Gamma}_{\mathrm{p}}=n_{\mathrm{p}}\left(\mu_{\mathrm{p}} \boldsymbol{E}\right)-D_{\mathrm{p}} \nabla n_{\mathrm{p}}
\end{equation}
\begin{equation}
   \label{Eq6}
\boldsymbol{\Gamma}_{\mathrm{n}}=-n_{\mathrm{n}}\left(\mu_{\mathrm{n}} \boldsymbol{E}\right)-D_{\mathrm{n}} \nabla n_{\mathrm{n}}
\end{equation}
, in which \( \mu_\text{e} \), \( \mu_\text{p} \) and \( \mu_\text{n} \) denote the mobilities of electrons, positive ions, and negative ions, respectively. \( D_\text{e} \), \( D_\text{p} \) and \( D_\text{n} \) denote the diffusion coefficients for electrons, positive ions, and negative ions, respectively. Specific expressions of them \cite{EN3602} are shown in TABLE \ref{tab:table1}.

The photoionization source term is described in more detail in Section \ref{2.2}. The reaction source terms in the continuity equation describe the effects of reaction processes (such as collision ionization, attachment and recombination) on the formation and consumption of charged species, which can be further expressed as: 
\begin{equation}
   \label{Eq7}
S_{\mathrm{R}, \mathrm{e}}=\alpha n_{\mathrm{e}}\left|\mu_{\mathrm{e}} \boldsymbol{E}\right|-\eta n_{\mathrm{e}}\left|\mu_{\mathrm{e}} \boldsymbol{E}\right|-\beta_{\mathrm{ep}} n_{\mathrm{e}} n_{\mathrm{p}}
\end{equation}

\begin{equation}
   \label{Eq8}
S_{\mathrm{R}, \mathrm{p}}=\alpha n_{\mathrm{e}}\left|\mu_{\mathrm{e}} \boldsymbol{E}\right|-\beta_{\mathrm{ep}} n_{\mathrm{e}} n_{\mathrm{p}}-\beta_{\mathrm{np}} n_{\mathrm{n}} n_{\mathrm{p}}
\end{equation}
\begin{equation}
   \label{Eq9}
S_{\mathrm{R}, \mathrm{n}}=\eta n_{\mathrm{e}}\left|\mu_{\mathrm{e}} \boldsymbol{E}\right|-\beta_{\mathrm{np}} n_{\mathrm{n}} n_{\mathrm{p}}
\end{equation}
, in which \(\alpha\) denotes the collision ionization coefficient, \(\eta\) denotes the attachment coefficient, \(\beta_{\mathrm{ep}}\) denotes the coefficient for electron-ion recombination, and \(\beta_{\mathrm{np}}\) denotes the coefficient for ion-ion recombination. The specific expressions of these coefficients \cite{EN3602} are shown in TABLE \ref{tab:table2}. 

\begin{table}[t]
\caption{\label{tab:table2} The expression of reaction coefficients, in which \(N\) is the number density of neutral species with a value of $2.45\times 10^{25}$ \(\text{m}^{-3}\) at 1 atm, and \(E/N\) is the reduced electric field with the unit in $V\cdot \text{m}^2$, \(p\) is the gas pressure with the unit in Pa}
\begin{ruledtabular}
\begin{tabular}{cc}
&expression\\
\hline
$\alpha[\text{m}^{-1}]$ & $\alpha=\left\{\begin{array}{l}
\left(3.4473 \times 10^{34}(E / N)^{2.985}\right) \cdot N \\
, E / N<4.6 \times 10^{-19} \\
\\
\left(11.269(E / N)^{1.159}\right) \cdot N \\
, E / N>4.6 \times 10^{-19}
\end{array}\right.$\\
\\
$\eta[\text{m}^{-1}]$& $\eta=\left\{\begin{array}{l}
\left(\left(2.0463 \times 10^{-20}-0.25379(E / N)\right.\right. \\
+1.4705 \times 10^{18}(E / N)^2 \\
\left.-3.0078 \times 10^{36}(E / N)^3\right) \cdot N \\
, 5 \times 10^{-20}<E / N<2 \times 10^{-19} \\
\\
\left(7 \times 10^{-21} \exp \left(-2.25 \times 10^{18}(E / N)\right)\right) \cdot N \\
, E / N>2 \times 10^{-19}
\end{array}\right.$\\
\\
$\beta_{\text{np}}[\text{m}^{3}\cdot\text{s}^{-1}]$ & $\beta_{\text{np}}=\left\{\begin{array}{l}
2 \times 10^{-13} p^{0.6336} \\
, 1<p<3.9 \times 10^4 \\
\\
2.28 \times 10^{-11} p^{-0.659} \\
, 3.9 \times 10^4<p<2.7 \times 10^5 \\
\\
6.867 \times 10^{-10} p^{-1.279} \\
, 2.7 \times 10^5<p<2 \times 10^6
\end{array}\right.$\\
\\
$\beta_{\text{ep}}[\text{m}^{3}\cdot\text{s}^{-1}]$ & $\beta_{\text{ep}}= 0$\footnote{The electron-ion recombination in SF\(_6\) at high pressure and low temperature is negligible, according to Refs.\cite{AGleizes_1989,PSpanel_1995}}\\

\end{tabular}
\end{ruledtabular}
\end{table}

The local field approximation is employed in this model. It should be acknowledged that, although the local field approximation can reduce numerical convergence difficulties, it neglects energy relaxation process, thereby inaccurately reflecting the dynamics of charged species near the computational boundaries \cite{zhu2021simulation}. The local energy approximation is more accurate. Previous research on SF\(_6\) streamer simulation \cite{10.1063/5.0223522} employed the local energy approximation and captured the accumulation of negative ions near protrusions. However, since the focus of this paper is on the dynamics inside the SF\(_6\) streamer channel, the local field approximation is sufficient to meet the research objectives.

The Poisson’s equation is used to calculate the distribution of electric potential \(V\): 
\begin{equation}
   \label{Eq10}
\nabla^2 V=-e\left(n_{\mathrm{p}}-n_{\mathrm{e}}-n_{\mathrm{n}}\right) / \varepsilon_0
\end{equation}
, where \(e\) is the elementary charge, and \(\epsilon_0\) is the vacuum permittivity. The electric field is calculated based on its defining equation:
\begin{equation}
   \label{Eq11}
\boldsymbol{E}=-\nabla V
\end{equation}

The finite element method (FEM) is used to solve the aforementioned equations. To improve the stability of numerical calculations, the streamline diffusion technique \cite{johnson1986streamline} is employed to mitigate the numerical instability caused by large numerical gradients.

\subsection{\label{2.2}Photoionization}
To date, a precise computational model for photoionization related to SF\(_6\) has not yet been established in fluid simulations. Levko \textit{et al.} \cite{10.1063/5.0131780} and Ou \textit{et al.} \cite{10.1063/5.0006140} ignored the effect of photoionization in the context of their research. Regarding simplified alternative approaches, Dhali \textit{et al.} \cite{14} and Luo \textit{et al.} \cite{reaction2} employed a uniform background ionization, while Sun \textit{et al.} \cite{10400497} and He \textit{et al.} \cite{9450688} added a constant source term to the continuity equation. However, these alternative approaches cannot capture the non-local effect of photoionization, which is crucial for supplying seed electrons at the front of the streamer and is essential for its propagation. 
 
 In this paper, Zhelezniak’s model \cite{ZK} for photoionization in air is employed to approximate the photoionization process in SF\(_6\), in order to reflect the non-local effects of the process. It is important to note that all parameters within this photoionization model are derived from air-related data, which inevitably introduces inaccuracies when quantitatively describing SF\(_6\). Consequently, this approach serves as a simplified alternative intended for the qualitative modeling of SF\(_6\) photoionization.
 
In Zhelezniak’s model, the photoionization source term \(S_{\mathrm{ph}}\) at the observation point \(\boldsymbol{r}\) due to source points emitting photoionizing UV photons at \(\boldsymbol{r}^{\prime}\) is defined as:
\begin{equation}
   \label{Eq12}
S_{\mathrm{ph}}(\boldsymbol{r})=\int_{V_1} \frac{I\left(\boldsymbol{r}^{\prime}\right) g\left(\boldsymbol{r}-\boldsymbol{r}^{\prime}\right)}{4 \pi\left|\boldsymbol{r}-\boldsymbol{r}^{\prime}\right|^2} \mathrm{~d} V_1
\end{equation}
, in which \(I\left(\boldsymbol{r}^{\prime}\right)\) is production of photons, \(g\left(\boldsymbol{r}-\boldsymbol{r}^{\prime}\right)\) is photo absorption function, by approximating the \(g\left(\boldsymbol{r}-\boldsymbol{r}^{\prime}\right)\), the integral model described by equation (\ref{Eq12}) can be turned into multiple Helmholtz equations (\ref{Eq13}) to simplify numerical computation \cite{Bourdon_2007,10.1063/1.2435934}:
\begin{equation}
   \label{Eq13}
S_{\mathrm{ph}}(\boldsymbol{r})=\sum_j S_{\mathrm{ph}}^j(\boldsymbol{r})
\end{equation}
with terms
\begin{equation}
   \label{Eq14}
\nabla^2 S_{\mathrm{ph}}^j(\boldsymbol{r})-\left(\lambda_j p_{\mathrm{O}_2}\right)^2 S_{\mathrm{ph}}^j(\boldsymbol{r})=-A_j p_{\mathrm{O}_2}^2 I(\boldsymbol{r})
\end{equation}
, in which \(p_{\mathrm{O}_2}\) denotes the partial pressure for oxygen, \(\lambda_j\) and \(A_j\) are parameters of Helmoltz equations. This paper employs Bourdon’s three-term expansion \((j=1,2,3)\) with parameter values consistent with those presented in Ref.\cite{Bourdon_2007}. It is noteworthy that the production of photons  \(I\left(\boldsymbol{r}\right)\) is directly proportional to the collision ionization production rate \(S_{\mathrm{i}}(\boldsymbol{r})\):
\begin{equation}
   \label{Eq15}
I(\boldsymbol{r})=\frac{p_{\mathrm{q}}}{p+p_{\mathrm{q}}} \xi \frac{v_u}{v_{\mathrm{i}}} S_{\mathrm{i}}(\boldsymbol{r})
\end{equation}
, in which \(S_{\mathrm{i}}(\boldsymbol{r})\) = \(\alpha n_{\mathrm{e}}\left|\mu_{\mathrm{e}} \boldsymbol{E}\right|\), \(p_{\mathrm{q}}\) = 30 Torr denotes the quenching pressure in air, \(p\) denotes the gas pressure, \(\xi\) denotes the photoionization efficiency, \(v_u\) denotes the excitation frequency, \(v_{\mathrm{i}}\) denotes the ionization frequency. 
The combined variable \(\xi \frac{v_u}{v_{\mathrm{i}}}\) is a weak function of the reduced electric field, taken as 0.06 \cite{Liuningyu}. Thus, for ambient air, the proportionality factor \(\frac{p_{\mathrm{q}}}{p+p_{\mathrm{q}}} \xi \frac{v_u}{v_{\mathrm{i}}}\) \(\approx\) 0.00228. In Section \ref{3}, it is assumed that \(I\left(\boldsymbol{r}\right)\) = 0.00228 \(S_{\mathrm{i}}(\boldsymbol{r})\), while in other sections, the proportionality factor \(\frac{p_{\mathrm{q}}}{p+p_{\mathrm{q}}} \xi \frac{v_u}{v_{\mathrm{i}}}\) is modified to reduce non-physical effects of photoionization.
\subsection{Boundary Condition and Initial Condition}

The boundary conditions in our model, including Boundary 1, Boundary 2, and Boundary 3 as shown in FIG.\ref{fig.geo}, are consistent with those specified in Ref.\cite{15}. The static voltage \( U_0 \) is applied to Boundary 1.

Initially, the number density of electrons and positive ions is assumed to follow a Gaussian distribution, expressed as:
\begin{equation}
   \label{Eq15}
 n_{\text{e(p)}} = n_{\text{max}} \exp\left(-\frac{(r - r_0)^2}{2s_0^2} - \frac{(z - z_0)^2}{2s_0^2}\right) 
\end{equation}
in which \( n_{\text{max}} \) denotes the peak density with a value of \( 10^{13} \, \text{m}^{-3} \),  \( (r_0, z_0) \) denotes the coordinates of the rod electrode tip, and the parameter \( s_0 \) is set to 0.1 mm.

The gas temperature \( T_\text{g} \) is set to 300 K, and the gas pressure \( p \) is maintained at \( 1.013 \times 10^5 \) Pa, consistent with a typical engineering scenario described in Ref.\cite{8035405}. These two parameters are assumed to remain constant throughout the entire streamer process.

\section{\label{3}Coherent Structures within SF\(_6\) Streamer Channel}
\subsection{Comprehensive Studies of Gas Gap Distance and Voltage Polarity}
\begin{figure}[h]
\centering
\includegraphics[width=8.5cm]{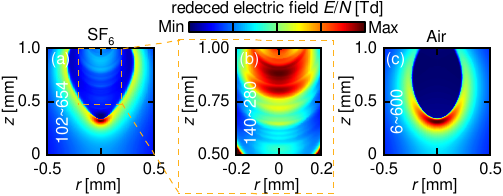}
\caption{\label{fig.compr}Evolution of the reduced electric field \(E/N\) for: (a) SF\(_6\) streamer; (b) local magnified view of SF\(_6\) streamer channel; (c) air streamer. Labels for \((E/N)_\text{min}\), \((E/N)_\text{max}\), and \(n_\text{e,max}\) are shown in each sub-figure, where \(n_\text{e,min}\) is fixed at 0.}
\end{figure}

As briefly mentioned in Section \ref{1}, the SF\(_6\) streamer channel exhibits different characteristics compared to air. A typical comparison between SF\(_6\) streamer and air streamer, as well as the local magnified view of the SF\(_6\) streamer channel, is shown in Fig.\ref{fig.compr}.

\begin{figure*}[t]
\centering
\includegraphics[width=17cm]{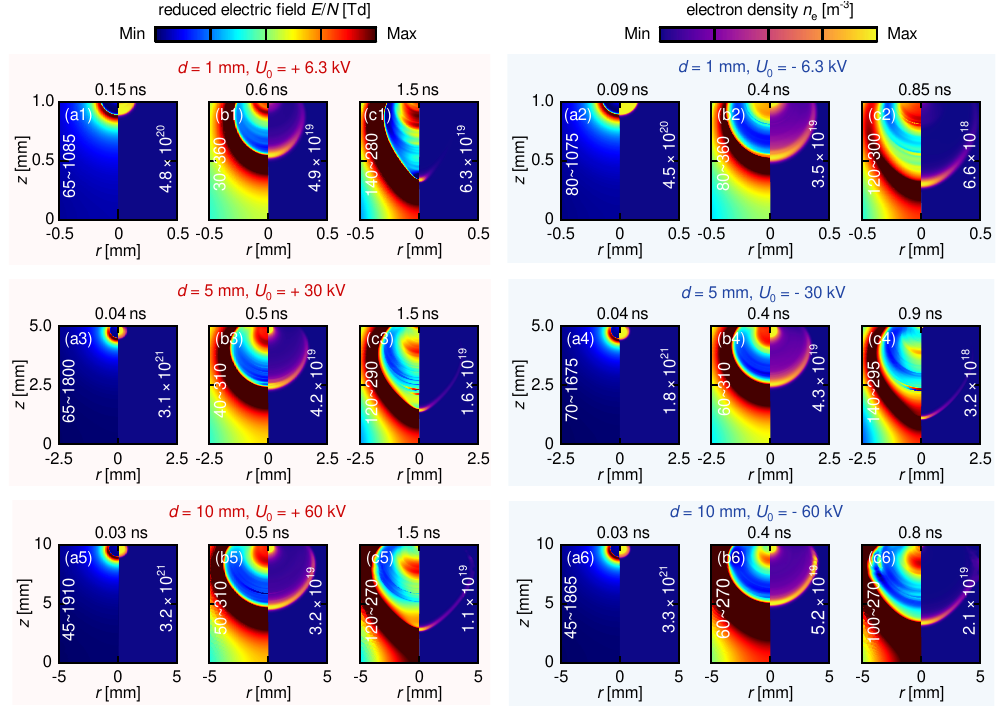}
\caption{\label{fig.normal0.002} Evolution of the reduced electric field \(E/N\) and electron density \(n_\text{e}\) under different conditions: (a1, b1, c1) \(d\) = 1 mm, \(U_0\) = +6.3 kV; (a2, b2, c2) \(d\) = 1 mm, \(U_0\) = -6.3 kV; (a3, b3, c3) \(d\) = 5 mm, \(U_0\) = +30 kV; (a4, b4, c4) \(d\) = 5 mm, \(U_0\) = -30 kV; (a5, b5, c5) \(d\) = 10 mm, \(U_0\) = +60 kV; (a6, b6, c6) \(d\) = 10 mm, \(U_0\) = -60 kV. The photon production \(I\left(\boldsymbol{r}\right)\) = 0.00228 \(S_{\mathrm{i}}(\boldsymbol{r})\) for all configurations. Labels for \((E/N)_\text{min}\), \((E/N)_\text{max}\), and \(n_\text{e,max}\) are shown in each sub-figure, where \(n_\text{e,min}\) is fixed at 0.}
\end{figure*}

The voltage polarity \cite{10.1063/1.4961925,Starikovskiy_2020} and gas gap distance \cite{RMorrow1997,Bujotzek_2015} significantly influence the dynamics of streamer discharge. To eliminate stochastic variations and gain insights into the general physics of the SF\(_6\) streamer channel, comprehensive studies focusing on these two factors are necessary. Accordingly, a set of simulation cases are performed with gas gap distance \(d\) of 1mm, 5mm, and 10mm. For each \(d\), streamer discharge is simulated under both positive and negative voltage polarities. The results are shown in FIG.\ref{fig.normal0.002}, where the minimum and maximum values of the color bar for the reduced electric field \(E/N\) have been adjusted to more clearly illustrate the field distribution within the streamer channel.

During the initial avalanche stage (not shown in any figures) of discharge , the electron density \(n_\text{e}\) reaches \( \sim 10^{21} \, \text{m}^{-3}\). This high \(n_\text{e}\) hinders the strong electronegativity of SF\(_6\) from manifesting during the early stage of the streamer, as shown in FIG.\ref{fig.normal0.002} (a1–a6). Specifically, although the streamer forms and effectively shields the internal electric field, reducing it to well below the critical value \((E/N)_{\text{cr}} = 360 \, \text{Td}\), the short formation time does not allow for the effective attachment of electrons that are generated during the avalanche stage. As a result, The SF\(_6\) streamer channel in the early stage remains a high-conductivity channel dominated by electrons, showing no distinct structural differences compared to that in air.\cite{wangzhen2022}.

As the streamer propagates, the electron density \(n_\text{e}\) inside the channel decreases to \(\sim 10^{18} \, \text{m}^{-3}\) due to attachment reactions, which have continuously taken place over the sub-nanosecond period, as shown in FIG.\ref{fig.normal0.002} (b1–b6). This creates a difference of \(\sim\) 1 order of magnitude in \(n_\text{e}\) between the streamer head and the channel, leading to an electron-deficient region. Additionally, the \(E/N\) inside the channel begins to recover, particularly at the streamer tail near the rod electrode, where it recovers to nearly twice the \((E/N)_\text{min}\). Although not very pronounced, minor spatial fluctuation in field strength can still be observed, predominantly in regions with lower \(n_\text{e}\), such as the channel interior and the area immediately behind the streamer head.

As the streamer propagates closer to the ground electrode, as shown in FIG.\ref{fig.normal0.002} (c1–c6), the electron density \(n_\text{e}\) inside the channel becomes significantly lower, by more than 3 orders of magnitude compared to that in the streamer head, further emphasizing the electron-deficient region. At this stage, the \(E/N\) within the channel exhibits continuous and pronounced spatial fluctuation in the same low \(n_\text{e}\) region as in the previous stage, but the fluctuation is significantly more pronounced than previously observed.

\begin{figure*}[t]
\centering
\includegraphics[width=17cm]{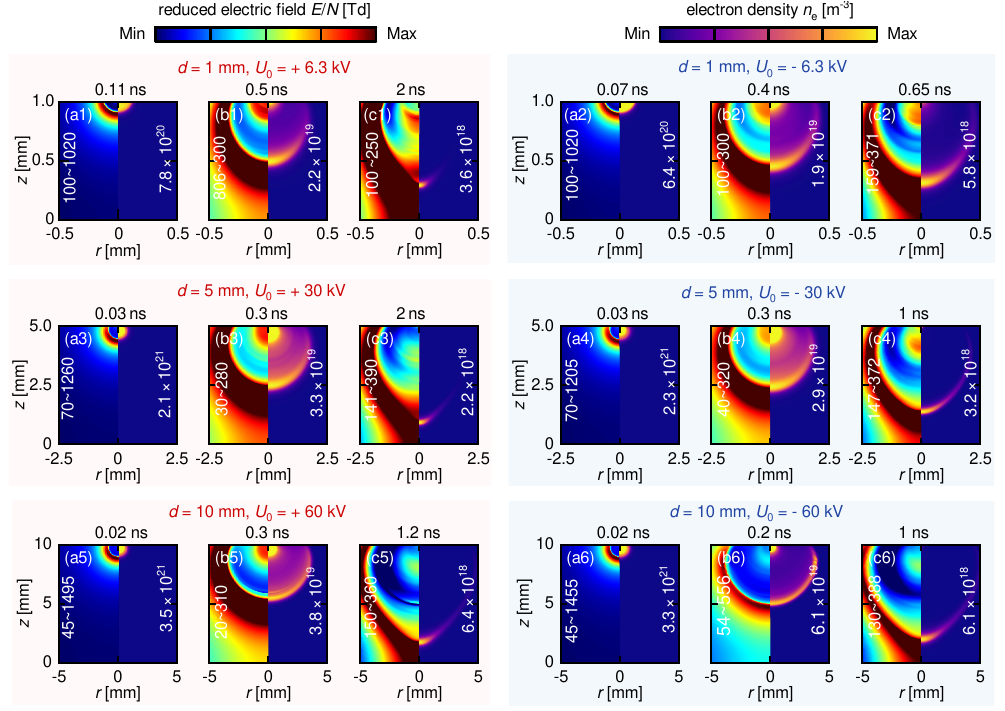}
\caption{\label{fig.normal0.1}Evolution of the reduced electric field \(E/N\) and electron density \(n_\text{e}\) under different conditions: (a1, b1, c1) \(d\) = 1 mm, \(U_0\) = +6.3 kV; (a2, b2, c2) \(d\) = 1 mm, \(U_0\) = -6.3 kV; (a3, b3, c3) \(d\) = 5 mm, \(U_0\) = +30 kV; (a4, b4, c4) \(d\) = 5 mm, \(U_0\) = -30 kV; (a5, b5, c5) \(d\) = 10 mm, \(U_0\) = +60 kV; (a6, b6, c6) \(d\) = 10 mm, \(U_0\) = -60 kV. The modified photon production \(I\left(\boldsymbol{r}\right)\) = 0.1 \(S_{\mathrm{i}}(\boldsymbol{r})\) for all configurations. Labels for \((E/N)_\text{min}\), \((E/N)_\text{max}\), and \(n_\text{e,max}\) are shown in each sub-figure, where \(n_\text{e,min}\) is fixed at 0.}
\end{figure*}

\subsection{General Phenomena within Streamer Channel: Coherent Structures}
In all the above configurations with varying \(d\) and \(U_0\), the SF\(_6\) streamer channel consistently exhibits general physical phenomena, characterized by coherent structures formed through nonlinear dynamics. These coherent structures include channel field recovery, electron-deficient region, and spatial fluctuation of the electric field (SFEF), none of which are typically observed in air \cite{Luque_2008}. Understanding these structures is critical for comprehending the fundamental processes of streamer discharge in SF\(_6\). Previous research has elucidated the mechanism of channel field recovery and electron-deficient region \cite{10.1063/5.0223522}. However, the SFEF phenomenon remains unexplained and seems unique to pure SF\(_6\), with no similar characteristics reported in SF\(_6\) mixtures \cite{PhysRevA.37.4396}, or C\(_4\)F\(_7\)N mixtures \cite{10.1063/5.0186055,Simonović_2024}. The physical validity and underlying mechanism of SFEF still remain unclear.

It can be observed that SFEF is a complex phenomenon, characterized by continuous fluctuations in field strength within the streamer channel, typically occurring in the low \(n_\text{e}\) regions. To validate its physical validity, it is essential to eliminate the impacts of model factors that may deviate from the real physical scenario. Therefore, potential factors that could introduce non-physical effects, such as photoionization intensity and mesh size, thoroughly examined in detail in Section \ref{4.1} and \ref{4.2}, respectively.

\section{\label{4}Validation of the physical validity of SFEF}

\subsection{\label{4.1}Impact of Photoionization Intnesity}

As detailed in Section \ref{2.2}, the parameters used in the photoionization model are based on air-related data, which introduces potential inaccuracies. It is crucial to assess whether the photoionization process could lead to SFEF as a non-physical phenomenon, a possibility that remains undetermined. Specifically, as discussed in Section \ref{3}, SFEF is typically observed in regions with low \(n_\text{e}\), where the photoionization process can provide a certain amount of electrons. This observation raises two key questions: 

(1) Is the occurrence of SFEF possibly due to insufficient electron replenishment caused by weak photoionization intensity?

(2) If the photoionization intensity is enhanced, can it prevent the occurrence of SFEF?
To address these questions, the production of photons  \(I\left(\boldsymbol{r}\right)\), which directly determines photoionization intensity, is modified to a level beyond realistic conditions. The rationale is that if SFEF still occurs under this modified level, the fundamental role of photoionization can be ruled out. In the modified model, the proportionality factor \(\frac{p_{\mathrm{q}}}{p+p_{\mathrm{q}}} \xi \frac{v_u}{v_{\mathrm{i}}}\) within  \(I\left(\boldsymbol{r}\right)\) is increased from 0.00228 to 0.1 (\(\sim\) 44 times that of air) for the cases where \(d\) = 1 mm and 5 mm. Consequently, the production of photons is expressed as \(I\left(\boldsymbol{r}\right)\) = 0.1 \(S_{\mathrm{i}}(\boldsymbol{r})\). It should be noted that, although our choice of the factor 0.1 is arbitrary, this relatively large value ensures an exaggerated modification as inferred from Refs.\cite{10.1063/1.1288407,10.1063/1.1743151,Janalizadeh_2019}. Thus it is intended purely for theoretical analysis of general physics rather than for quantitatively reflecting real physical scenarios.

SF\(_6\) streamer discharge is simulated with modified photoionization. The results are shown in FIG.\ref{fig.normal0.1}, where the minimum and maximum values of the color bar for the reduced electric field \(E/N\) have also been adjusted. During the early stage, as shown in FIG.\ref{fig.normal0.1} (a1-a6), the streamer channel exhibits the same properties as those observed in FIG.\ref{fig.normal0.002} (a1-a6), characterized by a high-conductivity channel with a relatively high \(n_\text{e}\). As the streamer propagates over a longer period, as shown in FIG.\ref{fig.normal0.1} (b1-b6), the characteristics of the streamer channel remain consistent with those in FIG.\ref{fig.normal0.002} (b1-b6), displaying the electron deficient region and the presence of SFEF. The enhancement of photoionization does not result in any fundamental changes in the coherent structures. However, it should be noted that, in some cases—such as those shown in FIG.\ref{fig.normal0.1} (b3), (b4) and (b5)—the \(n_\text{e}\) within the streamer channel increases to \(\sim 10^{19} \, \text{m}^{-3}\), reducing the difference from the streamer head to less than half an order of magnitude. This makes the electron-deficient region less pronounced, and it can also be observed that the SFEF in these cases becomes less prominent.  This observation further supports the potential correlation between \(n_\text{e}\) and SFEF. Finally, as the streamer propagates closer to the ground electrode, as shown in FIG.\ref{fig.normal0.1} (c1-c6), a pronounced electron deficient region and SFEF are observed within the streamer channel, indicating that even with exaggeratedly enhanced photoionization intensity, the occurrence of SFEF cannot be prevented.

All the above results indicate that, despite the exaggerated enhancement of photoionization intensity, the field strength within the SF\(_6\) streamer channel remains consistent with the spatial fluctuation patterns observed in FIG.\ref{fig.normal0.002}. This finding suggests that the occurrence of SFEF is independent of variations in photoionization intensity, indicating that other factors may be responsible for governing it.

\subsection{\label{4.2}Impact of Mesh Size}

\begin{figure}[t]
\centering
\includegraphics[width=8.5cm]{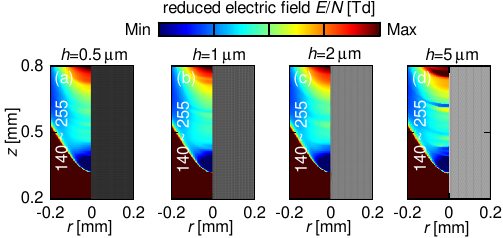}
\caption{\label{fig.mesh} Distribution of mesh and reduced electric field \(E/N\) for different mesh size \(h\): (a) \(h = 0.5 \, \mu \text{m}\), (b) \(h = 2 \, \mu \text{m}\), (c) \(h = 1 \, \mu \text{m}\), and (d) \(h = 5 \, \mu \text{m}\). Labels for \((E/N)_\text{min}\), \((E/N)_\text{max}\) are shown in each sub-figure.}
\end{figure}

In the FEM simulation of SF\(_6\) streamer discharge, accurately resolving regions with steep electron density gradients requires a sufficiently fine mesh \cite{Bessières_2007}, otherwise, non-physical fluctuation resulting from numerical errors may occur \cite{OKUDA1972475}. Specifically, to effectively capture charge separation, the maximum mesh size should be smaller than the Debye length \(\lambda_\text{D} = \sqrt{\frac{\epsilon_0 k_\text{B} T_\text{e}}{n_\text{e} e^2}}\), in which the electron temperature \( T_\text{e} \) \(\approx\) 4 eV and electron density \( n_\text{e} \) \(\approx\) \( 10^{19} \) m\(^{-3}\). Additionally, the minimum mesh size should adhere to certain constraints to ensure physical accuracy. First, \(h\) should be larger than the electron displacement within a single time step (CFL condition \cite{TEUNISSEN2018156}), given by \( v_\text{e} \cdot \Delta t \), in which electron velocity \( v_\text{e} \) \(\approx\) \( 2 \times 10^5 \) m/s and calculation time step \( \Delta t \) \(\approx\) \( 10^{-12} \) s. Second, \(h\) should also be larger than the mean free path of electrons, given by \(\frac{1}{N \sigma}\), in which number density of neutral species \( N \) \(\approx\)  \( 2.45 \times 10^{25} \) m\(^{-3}\) and effective collision cross section \( \sigma \) \(\approx\)  \( 10^{-19} \) m\(^2\). Therefore, the mesh size should satisfy the range 0.42 µm < \(h\) < 5.25 µm.

To investigate whether the mesh size \( h \) determines the occurrence of SFEF within the streamer channel, the mesh size is adjusted within the specified range. For simplicity, simulations are performed only under the configuration of \( d = 1 \) mm and \( U_0 = + 6.3 \) kV. Within the streamer channel region (0 < \(r\) < 0.2 mm, 0.2 mm < \(z\) < 0.8 mm), \( h \) is varied, taking values of 0.5 µm, 1 µm, 2 µm, and 5 µm, respectively. The results are shown in FIG.\ref{fig.mesh}, where the minimum and maximum values of the color bar for the reduced electric field \(E/N\) have also been adjusted. It can be observed that SFEF occurs across all mesh size cases, even at the minimum \(h\) = 0.5 µm. This finding suggests that the occurrence of SFEF is independent of the mesh resolution employed in the simulations, thereby ruling out the possibility that it is a non-physical phenomenon caused by the mesh size.

In summary, the two potential factors (photoionization intensity and mesh size) do not fundamentally determine the occurrence of SFEF. This suggests that the fluctuation is a general coherent structure within SF\(_6\) streamer channel. The following sections further investigate its underlying physical mechanism through comparative analysis.

\section{\label{5}Search for the physical mechanism of SFEF}
\subsection{\label{5.1}Establishing a Comparative Basis via Voltage Adjustment}

\begin{figure*}[t]
\centering
\includegraphics[width=17cm]{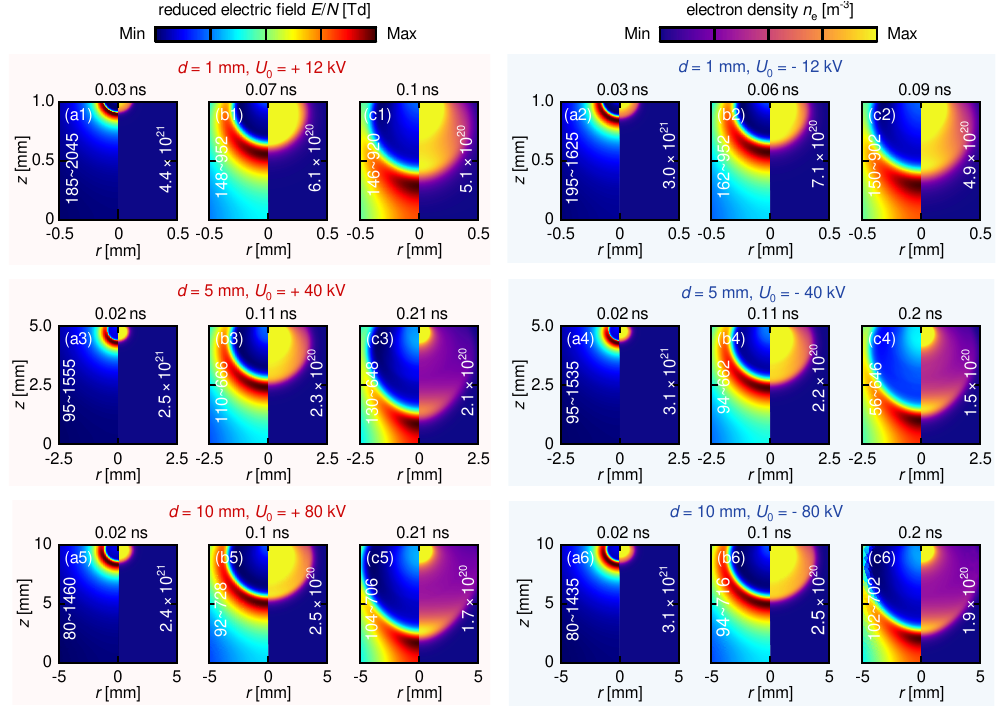}
\caption{\label{fig.over0.1}Evolution of the reduced electric field \(E/N\) and electron density \(n_\text{e}\) under different conditions: (a1, b1, c1) \(d\) = 1 mm, \(U_0\) = +12 kV; (a2, b2, c2) \(d\) = 1 mm, \(U_0\) = -12 kV; (a3, b3, c3) \(d\) = 5 mm, \(U_0\) = +40 kV; (a4, b4, c4) \(d\) = 5 mm, \(U_0\) = -40 kV; (a5, b5, c5) \(d\) = 10 mm, \(U_0\) = +80 kV; (a6, b6, c6) \(d\) = 10 mm, \(U_0\) = -80 kV. The modified photon production \(I\left(\boldsymbol{r}\right)\) = 0.1 \(S_{\mathrm{i}}(\boldsymbol{r})\) for all configurations. Labels for \((E/N)_\text{min}\), \((E/N)_\text{max}\), and \(n_\text{e,max}\) are shown in each sub-figure, where \(n_\text{e,min}\) is fixed at 0.}
\end{figure*}

To investigate the physical mechanism behind SFEF, a direct approach is to compare scenarios where such fluctuation is and is not pronounced. Based on the findings of Section \ref{3}, it is evident that SFEF typically occurs in conjunction with the electron deficient region, suggesting a potential correlation between SFEF and \(n_\text{e}\) from a phenomenological perspective. This observation leads to the following hypothesis: by artificially increasing the \( n_\text{e} \) within the streamer channel, thereby reducing the prominence of electron-deficient areas, it might be possible to also diminish the prominence of SFEF, providing a basis for comparative analysis

To explore this hypothesis, the applied voltage \( U_0 \) is chosen as the adjustable parameter to artificially increase \( n_\text{e} \), based on the theoretical premise that \( U_0 \) directly determines the rate of effective ionization. Consequently, \(U_0\) is increased to overvoltage levels of ±12kV, ± 40 kV, and ± 80 kV, corresponding to \(d\) values of 1mm, 5mm, and 10mm, respectively, significantly exceeding the static breakdown threshold. However, in practice, streamer discharge will inevitably occur at a lower voltage, before \(U_0\) reaches the overvoltage level. Thus it should be noted that the overvoltage discharge process simulated here would not occur under real static conditions; it represents an idealized scenario chosen solely to provide a theoretical basis for comparison. 

The discharge process under overvoltage conditions is shown in FIG.\ref{fig.over0.1}. At all stages of the discharge, the electron density \(n_e\) within the streamer channel exceeds \(10^{19}\) m\(^{-3}\). As the streamer propagates from the rod electrode toward the ground electrode, \(n_e\) is significantly higher compared to that under normal voltage condition (FIG.\ref{fig.normal0.002} and FIG.\ref{fig.normal0.1}). Additionally, the field strength within the streamer channel shows no significant fluctuation at any stage of the discharge, indicating that SFEF is not pronounced under overvoltage condition. This finding provides a basis for further investigation into the physical mechanism behind SFEF.

\subsection{\label{5.2}Physical Mechanism of SFEF}
The electric field strength, as calculated by combining Poisson's equation (\ref{Eq10}) and the defining equation (\ref{Eq11}), is fundamentally determined by the space charge distribution. Therefore, to understand the physical mechanism behind SFEF, it is essential to examine both the spatial distribution of space charge and the distribution of individual charged species.

The space charge density \(\rho_\text{v}\), which quantifies the amount of net charge, is expressed as:
 \begin{equation}
   \label{Eqn}
\rho_\text{v} = e(n_\text{p} - n_\text{e} - n_\text{n})
\end{equation}
, and the spatial distribution of \(\rho_\text{v}\), \(n_\text{e}\), \(n_\text{n}\), and \(n_\text{p}\) along the axis (\(r\) = 0) at normal voltage and overvoltage levels are plotted in FIG.\ref{fig.1Dnormal} and FIG.\ref{fig.1Dover}, respectively. By comparing the results at two voltage levels, the following inferences can be made:
\begin{figure}[t]
\centering
\includegraphics[width=8.5cm]{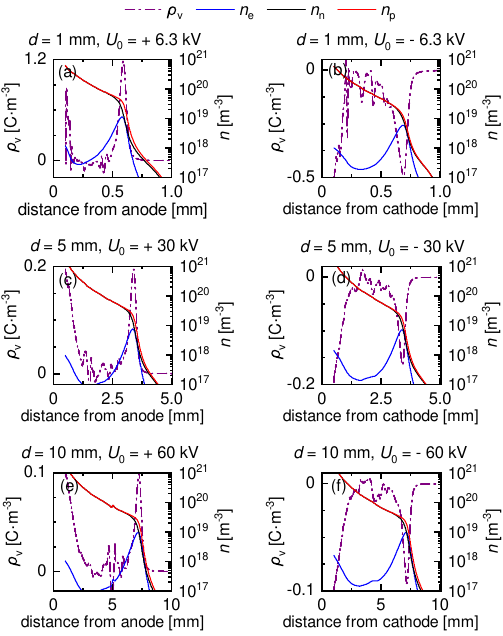}
\caption{\label{fig.1Dnormal} Distribution of the space charge density \(\rho_\text{v}\), electron density \(n_\text{e}\), negative ion density \(n_\text{n}\) and positive ion density \(n_\text{p}\) along the axis (\(r=0\)), illustrating the correlation between charged species and spatial fluctuation. All configurations are under normal voltage level: (a) \(d\) = 1 mm, \(U_0\) = +6.3 kV; (b) \(d\) = 1 mm, \(U_0\) = -6.3 kV; (c) \(d\) = 5 mm, \(U_0\) = +30 kV; (d) \(d\) = 5 mm, \(U_0\) = -30 kV; (e) \(d\) = 10 mm, \(U_0\) = +60 kV; (f) \(d\) = 10 mm, \(U_0\) = -60 kV. }
\end{figure}

\begin{figure}[t]
\centering
\includegraphics[width=8.5cm]{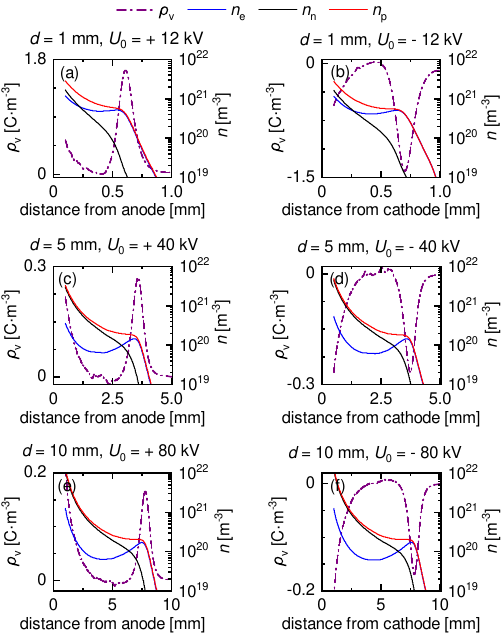}
\caption{\label{fig.1Dover}Distribution of the space charge density \(\rho_\text{v}\), electron density \(n_\text{e}\), negative ion density \(n_\text{n}\) and positive ion density \(n_\text{p}\) along the axis (\(r=0\)), as a comparative basis. All configurations are under overvoltage level: (a) \(d\) = 1 mm, \(U_0\) = +12 kV; (b) \(d\) = 1 mm, \(U_0\) = -12 kV; (c) \(d\) = 5 mm, \(U_0\) = +40 kV; (d) \(d\) = 5 mm, \(U_0\) = -40 kV; (e) \(d\) = 10 mm, \(U_0\) = +80 kV; (f) \(d\) = 10 mm, \(U_0\) = -80 kV. }
\end{figure}

(1) Under normal voltage condition, \(n_\text{e}\) within the streamer channel is \(\sim\)2 orders of magnitude lower than that at the streamer head, resulting in an electron deficient region. Consequently, the space charge within the streamer channel is primarily composed of ions rather than electrons, indicating that the streamer channel functions as an ion-conducting channel. However, under overvoltage condition, the situation changes dramatically; the difference in \(n_\text{e}\) between the streamer head and the channel is less than half an order of magnitude.  As a result, the presence of a large number of free electrons within the streamer channel causes it to function as an electron-conducting channel.

(2) In all simulation cases, the maximum drift velocity of ions within the channel, as derived from numerical calculations, is \(\sim\)\(10^2\) m/s, which is significantly lower than that of electrons, at \(\sim\)\(10^5\) m/s. This roughly 1000-fold disparity leads to a significant difference in how quickly charged species within the streamer channel respond to local charge relaxation. Specifically, for ions, within the simulation time of \(\sim\)ns, the maximum displacement is \(\sim\)0.1 \(\mu\)m, which is much smaller than the spatial scale of a single fluctuation (peak-to-peak distance, \(\sim\)10 \(\mu\)m) as shown in FIG.\ref{fig.1Dnormal}. Moreover, 0.1 \(\mu\)m is smaller than the minimum mesh size \(h\) specified in Section \ref{4.2}, making it difficult to capture accurately in numerical calculations. These factors result in the ions inside the channel being unable to respond promptly to local charge relaxation, similar to the behavior of surface charge in solid-state dielectrics \cite{10.1063/1.5096228}. In contrast, the situation is quite the opposite for electrons. Within the simulation time of \(\sim\)ns, the maximum displacement of electrons is \(\sim\)100 \(\mu\)m, which is sufficient to respond promptly to any possible charge fluctuation in space. Furthermore, it satisfies the mesh size constraints, allowing the response process to be accurately calculated. Additionally, the motion of ions is hindered by ion recombination reactions, whereas electron recombination is neglected in the model (as discussed in Table \ref{tab:table2}), further amplifying the differences between them.

(3)	Under normal voltage condition, once charge separation occurs within the ion-conducting streamer channel, it exhibits strong local characteristics dominated by ions. Specifically, this separation is not easily neutralized by ions outside the local area, leading to fluctuation in the space charge and, consequently, in the channel field strength. This behavior is similar to certain local surface charge patterns observed in solid-state dielectrics \cite{NP,NC}. Under overvoltage condition, the abundant free electrons within the electron-conducting streamer channel, can rapidly respond to local fluctuation in the space charge, migrating to regions of charge separation to neutralize them. This action suppresses the space charge fluctuation within the channel, making the fluctuation of field strength not pronounced.

In summary, the essence of SFEF lies in space charge fluctuation, driven by the local characteristics of charge separation within the ion-conducting channel. This local characteristics make it difficult for charge separation to be neutralized. However, the above analysis is based on the assumption that some charge separation has already occurred within the streamer channel. Therefore, it remains necessary to further investigate the initiation of charge separation.

\subsection{\label{5.3}Initiation of Charge Separation}
\begin{figure}[t]
\centering
\includegraphics[width=8.5cm]{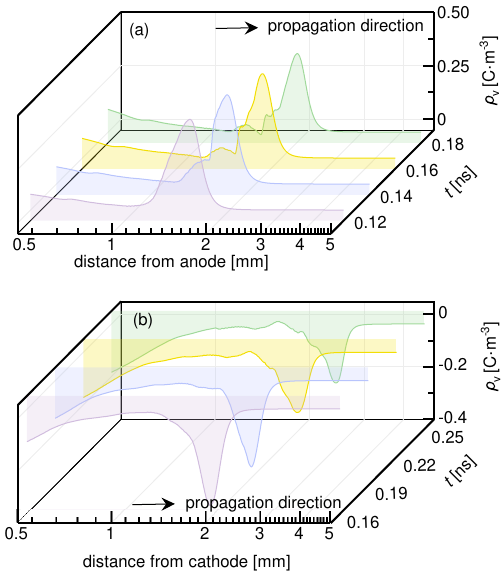}
\caption{\label{fig.1Dini} Evolution of the space charge density \(\rho_\text{v}\) along the axis (\(r=0\)), illustrating the initiation of charge separation. (a) The case with \(d = 5 \, \text{mm}\) and \(U_0 = +30 \, \text{kV}\), corresponding to the positive streamer. (b) The case with \(d = 5 \, \text{mm}\) and \(U_0 = -30 \, \text{kV}\), corresponding to the negative streamer. The propagation direction of the streamer is highlighted in both figures.}
\end{figure}

The typical cases, with \(U_0\) = ±30 kV and \(d\) = 5 mm, are selected to analyze the initiation of charge separation. The space charge distributions along the axis (\(r=0\)) for positive and negative streamers are shown in FIG.\ref{fig.1Dini} (a) and (b) respectively. Each case includes the time periods covering the entire process of space charge separation initiation.

\begin{figure}[t]
\centering
\includegraphics[width=8.5cm]{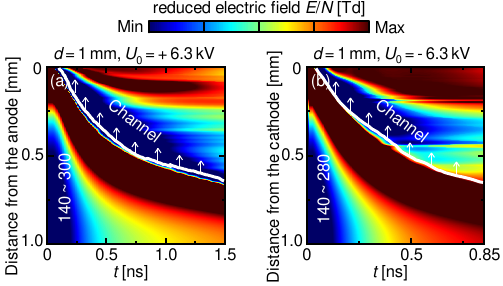}
\caption{\label{fig.2Dtime} Spatiotemporal plots of the reduced electric field \(E/N\) along the axis (\(r=0\)), illustrating the evolution of SFEF. (a) The case with \(d = 1 \, \text{mm}\) and \(U_0 = +6.3 \, \text{kV}\), corresponding to the positive streamer. (b) The case with \(d = 1 \, \text{mm}\) and \(U_0 = -6.3 \, \text{kV}\), corresponding to the negative streamer. The white lines show the channel edges and the arrows point towards the channel interior.}
\end{figure}

It is evident that, regardless of the polarity of the streamer, charge separation consistently occurs at the rear edge of the streamer head (which also corresponds to the rear edge of the ionization wave). This occurs because the electric field shielding effect generated by the streamer head is strongest at this location, leading to the highest rate of attachment reactions.  At this point, electrons are attached in large quantities, forming negative ions. Once electrons transform into negative ions, the drift velocity of the charged species decreases significantly, resulting in the accumulation of negative ions and subsequently altering the local space charge distribution.

As for the positive streamer, this accumulation locally reduces the positive space charge at the rear edge of the streamer head as shown in FIG.\ref{fig.1Dini} (a). As for the negative streamer, it locally increases the negative space charge at the same location as shown in FIG.\ref{fig.1Dini} (b). The accumulation of negative ions and the resultant change in local space charge lead to the initiation of space charge separation at the rear edge of the streamer head. As the streamer head propagates forward, the ion-dominated charge separation region is passively carried into the streamer channel, preserving its local characteristics and ultimately resulting in SFEF. Additionally, as the ionization of the streamer head progresses to new positions ahead, new charge separation continues to occur at the rear edge of the streamer head. However, it is important to note that once charge separation is carried into the channel, it influences the local electric field at the rear edge of the streamer head, which, in turn, affects the generation of new charge separation. This influence varies between positive and negative streamers, resulting in the continuous form of the SFEF and exhibiting polar effect, as shown in FIG.\ref{fig.2Dtime}. This polarity effect should be further investigated in the future.

\section{\label{6}effect of electric field uniformity}

\begin{figure}[b]
\centering
\includegraphics[width=8.5cm]{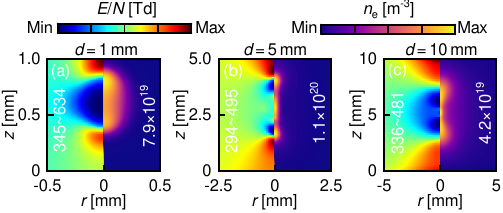}
\caption{\label{fig.2Duniform}Evolution of the reduced electric field \(E/N\) and electron density \(n_\text{e}\) of double-headed streamer in uniform fields: (a) \(d\) = 1 mm; (b) \(d\) = 5 mm; (c) \(d\) = 10 mm. Labels for \((E/N)_\text{min}\), \((E/N)_\text{max}\), and \(n_\text{e,max}\) are shown in each sub-figure, where \(n_\text{e,min}\) is fixed at 0.}
\end{figure}

To determine if SFEF is a unique phenomenon to non-uniform electric fields, we conduct simulations of SF\(_6\) streamer discharge in uniform electric fields. The electrode geometry is modified to a parallel plate configuration in 2D axisymmetric coordinate, similar to the setup described in Ref. \cite{https://doi.org/10.1029/2012GL051088}, ensuring a uniformly distributed background electric field. The initial charge species are placed at the center of the gas gap to initiate a double-headed streamer, allowing simultaneous observation of both positive and negative streamers during a single discharge event. The gas gap distance \(d\) is varied to 1 mm, 5 mm, and 10 mm, and the results are presented in Fig. \ref{fig.2Duniform}.

The simulation results indicate that SFEF does not manifest in uniform fields. In uniform fields, the initial electric field remains consistently distributed across the entire region. When the field strength at any point reaches the streamer threshold, the entire region exceeds the threshold, resulting in a relatively high background field. The effect of a uniform electric field scenario resembles that of an overvoltage condition, where a high ionization level is maintained within the streamer channel. Consequently, a substantial number of free electrons are generated, which suppresses the conditions necessary for SFEF formation. Therefore, the occurrence of SFEF is strongly dependent on the non-uniformity of the electric field. SFEF is observed exclusively in highly non-uniform fields and does not occur in uniform fields.

\section{Conclusions}
In this paper, we have simulated and analyzed the SF\(_6\) streamer channel in highly non-uniform fields. To comprehensively investigate the dynamics of streamer channel, we perform simulations under two key discharge parameters (\(d\) and \(U_0\)). The results indicate that the streamer channel in SF\(_6\) can exhibit several general coherent structures distinct from those in air, including the electron-deficient region, channel field strength recovery, and SFEF,  among which SFEF remains relatively less understood. The paper mainly focuses on an in-depth and thorough investigation of SFEF, validating its physical validity and revealing its potential physical mechanism.

(1) Although the photoionization intensity and mesh size in the model could potentially introduce non-physical effects, modifications to these factors indicate that they do not fundamentally determine the occurrence of SFEF. This finding rules out the possibility of SFEF being a non-physical phenomenon and confirms its physical validity.

(2) Comparative analysis indicates that the physical mechanism underlying SFEF is based on space charge fluctuation, driven by the local characteristics of charge separation within the ion-conducting channel. These local characteristics are governed by the low drift velocity of ions, making it difficult for the charge separation to be neutralized, which ultimately leads to SFEF.

(3) Charge separation is a precondition for SFEF; therefore, the initiation process of charge separation is further examined. It originates from the accumulation of negative ions at the rear edge of the streamer head. As the streamer head propagates, this charge separation continues to occur and is passively carried into the channel, causing SFEF to manifest in a continuous form.

(4) The occurrence of SFEF strongly dependent on the non-uniformity of the electric field, occurring exclusively in highly non-uniform fields and not in uniform fields.

The findings of this paper provide an in-depth understanding for the general coherent structure, SFEF, within the SF\(_6\) streamer channel in highly non-uniform fields. The fluctuation within the streamer channel will inevitably have a potential impact on the propagation of the streamer. Consequently, these results suggest a potential avenue for further exploration into the physical mechanisms governing the nonlinear breakdown voltage of SF\(_6\) and for optimizing the insulation strength of gas-insulated electrical equipment.

\section*{Acknowledgment}
The authors gratefully acknowledge the funding support from the National Natural Science Foundation of China (Contract No.~52277154). 
\section*{AUTHOR DECLARATIONS}
Conflict of Interest. The authors have no conflicts to disclose.
\section*{DATA AVAILABILITY}
The data that support the findings of this study are available from the corresponding author upon reasonable request.

\Large
\bibliography{references}

\end{document}